\documentstyle[amssymb,aps]{revtex}

\begin{document}
\draft
\title{A Theory of Phonon Spectroscopy in the Fractional Quantum Hall Regime}
\author{Keith A. Benedict, R.K. Hills and C.J. Mellor}
\address{School of Physics, University of Nottingham, \\
University Park, Nottingham, NG7 2RD U.K.}
\date{\today }
\maketitle

\begin{abstract}
We describe a theoretical framework for the interpretation of time-resolved
phonon absorption experiments carried out in the fractional quantum Hall
regime of a magnetically quantized two-dimensional electron system (2des).
The only phonons which can be absorbed at low temperature are those whose
energies exceed the magnetoroton gap predicted by Girvin, MacDonald and
Platzman. The rate of energy transfer from the phonons to the electron
liquid is entirely controlled by the creation of these collective
excitations. Using simple isotropic approximations for the phonon
propagation and electron-phonon coupling we obtain analytic results for the
regime in which the electron temperature and the characteristic temperature
of the phonons are much less than the gap and identify the way in which the
dispersion curve of the magnetorotons could be extracted from time and angle
resolved experiments.
\end{abstract}

\section{Introduction\label{Intro}}

Most of what we know experimentally about two-dimensional electron systems
has been gleaned from the analysis of D.C. transport measurements.
Typically, a constant electric current is passed though a 2d system while
the longitudinal and Hall voltages are monitored as the temperature,
magnetic field or density are varied. Recently there has been a move toward
the development of complementary techniques for acquiring information about
the novel states of electronic matter that arise when these 2d systems are
subject to strong magnetic fields. One such technique is phonon absorption
spectroscopy\cite{gen-phonon,fqh-phonon,ep2ds}. A pulse of non-equilibrium
acoustic phonons is injected into the substrate of a semiconductor sample
which has a 2d electron system formed at a heterojunction. The phonons
travel ballistically across the substrate until they encounter the sheet of
electrons where a small number of them are absorbed. Acoustic phonons are
particularly well suited to the study of 2d electron systems since they have
energies and wave vectors that are well matched to the low energy collective
excitations of these systems. Measurements of the rate at which energy is
absorbed by the electrons therefore contain important spectroscopic
information about the states of the 2d electron system.

The aim of this paper is to develop a framework for the interpretation of
phonon absorption experiments carried out in the fractional quantum Hall
regime. These experiments are now at a sufficiently sophisticated stage that
both time\cite{ep2ds} and, more recently, angle resolved measurements\cite
{Nijmegen} can be made. This paper will focus on our basic picture of what
is happening in the experiments. The calculations presented will be entirely
analytic, based on asymptotic analyses of general formulae making a wealth
of simplifying approximations to expose the qualitative features to be
expected. A more detailed numerical analysis which avoids some of the cruder
approximations is currently in progress and will be published as a sequel to
this work.

The structure of this paper is as follows. In section 2 an outline of the
procedure employed in the phonon absorption experiments will be given. In
section 3 the pertinent features of electrons and phonons in semiconductor
materials will be briefly reviewed while section 4 will review the standard
picture of the fractional quantum Hall state. We hope that this paper will
be of interest to both the community which studies the fractional quantum
Hall effect and the community which studies phonons: sections 3 and 4 are
intended to introduce the relevant ideas and to fix our notation. Section 5
will discuss the phonon transport aspects of the experiments. In section 6
the basic formulae for the rate of energy transfer from phonons to the
electron liquid in the fractional quantum Hall state will be introduced and
the kinematics of the absorption discussed. In section 7 asymptotic
expressions for the energy transfer rate will be derived employing an
isotropic approximation for the phonon properties. In section 8 the time
dependent aspects of the experiments are discussed. In section 9 the
processes leading to equilibrium within the 2DES are discussed qualitatively
(a more quantitative discussion of some aspects of this matter will appear
elsewhere). Section 10 contains a summary of our picture of what happens in
the phonon absorption experiments and what information could be extracted
from them. Various technical details of the calculations are relegated to
appendices.

Throughout this paper we will be discussing phonons characterized by
three-dimensional wave-vectors and collective excitations of the electronic
system characterized by two-dimensional wave vectors. In order to
distinguish these we will use underlined characters, $\underline{Q}$, to
represent 3d vectors and bold characters, ${\bf q}$, to represent 2d
vectors. Where a 2d vector arises as the projection of a 3d vector onto the $%
xy$-plane (see inset to figure \ref{schematic}) this will be denoted by $%
\underline{Q}=\left( {\bf q},q_{z}\right) $.

\section{Experimental Details\label{Experimental}}

An ideal phonon spectroscopy experiment would employ a monochromatic source
of phonons on one side of a device and a tunable phonon detector on the
opposite side with the 2d electron system lying between them. The fraction
of phonons absorbed from each mode could then be measured. In practice there
are no sources of monochromatic phonons which operate in high magnetic
fields and, similarly, no spectrally sensitive phonon detectors. The best
that one can hope to do at present is to inject a pulse of phonons with a
distribution of modes and detect it on the far side of the 2DES with a
bolometer. In fact, even this has yet to be done. What is actually done at
present involves the use of the 2DES itself as a bolometer.

A schematic picture of the samples used in the experiment is shown in figure
1. The basic device consists of a GaAs/AlGaAs heterojunction grown on top of
a thick ($\sim 2\mathop{\rm mm}$) substrate of semi-insulating GaAs. The
active area of 2DES is defined lithographically together with long
connections to heavily doped (3d) contacts. The active area is patterned
into a long meander-line of 2DES so that electrically it is a long, thin,
Hall bar with an aspect ratio of $\sim 30:1$. This patterning increases the
sensitivity of the device as a bolometer by increasing the longitudinal
resistance. To compensate for this, exceptional care is taken to reduce the
capacitance of the sample and the leads with respect to the ground so that
the RC time constant of the measuring circuit is not too large \cite{ep2ds}.
Small area metallic films are evaporated onto the back face of the device
and separate pairs of contacts made to each of these to allow them to act as
heaters. The application of a current pulse to a heater leads to Joule
heating within the film. Phonons are emitted from the film into the GaAs
with a black-body distribution of energies characterized by a pulse
temperature, $T_{\phi }$, which can be accurately predicted from the power
dissipated in the film using acoustic mismatch theory\cite{mismatch}.

Acoustic phonons in a semiconductor material such as GaAs travel
ballistically in the bulk of the material for frequencies up to $\sim 1%
\mathop{\rm THz}$ (this limit is set by the onset of Rayleigh scattering
from isotopic impurities) which is much higher than the frequencies of
interest here. The constant frequency surfaces in $\underline{Q}$-space are
significantly non-spherical so that the phonon group and phase velocities
are markedly different except along high symmetry directions. In fact the
curvature of the constant frequency surfaces changes sign so that phonon
wavepackets with a range of $Q$-vectors tend to be focussed into preferred
directions\cite{focus}. Another consequence of the elastic anisotropy of the
semiconductors is that phonon modes are only strictly transverse or
longitudinal along high symmetry directions.

Throughout the experiment a constant current is passed through the 2d
electron system and the two-terminal voltage monitored. The experiments are
conducted at the centre of the quantum Hall plateau at $\nu =1/m$. At zero
temperature the two terminal resistance would be identical to the quantized
Hall resistance $R_{2}=1/\overline{\sigma _{H}}$, $\overline{\sigma _{H}}%
=e^{2}/mh$. At a finite temperature the resistance is 
\begin{eqnarray*}
R_{2}\left( T_{e}\right) &\sim &\frac{\overline{\sigma _{H}}+n\delta \sigma
_{xx}\left( T_{e}\right) +\delta \sigma _{xy}\left( T_{e}\right) }{\left( 
\overline{\sigma _{H}}+n\delta \sigma _{xx}\left( T_{e}\right) \right)
^{2}+\left( \delta \sigma _{xy}\left( T_{e}\right) \right) ^{2}} \\
&\sim &\frac{1}{\overline{\sigma _{H}}}\left( 1+\frac{n\delta \sigma
_{xx}\left( T_{e}\right) }{\overline{\sigma _{H}}}-\frac{\delta \sigma
_{xy}\left( T_{e}\right) }{\overline{\sigma _{H}}}\right)
\end{eqnarray*}
where $\delta \sigma _{\mu \nu }\left( T_{e}\right) $ is the finite
temperature correction to quantized conductance tensor for a square 2des and 
$n$ is the number of squares in series in the device. One expects that at
the plateau centre $\delta \sigma _{xy}\left( T_{e}\right) $ will be small
and hence, given that $n\sim 30$, the change in the two-terminal resistance
is dominated by $\delta \sigma _{xx}\left( T_{e}\right) $.

Before the application of the pulse to the heater, the voltage across the
2DES is constant and has the value characteristic of the base temperature, $%
T_{0}$, of the system. When the current is applied to the heater, a pulse of
phonons is injected into the substrate which travels ballistically across
until it hits the 2DES, only when this occurs does the voltage change. The
substrate is sufficiently thick that the component of the phonon pulse which
contains the fast ``LA'' modes spatiotemporally separates from the slower
component containing the ``TA'' modes. Some of the phonons from the pulse(s)
are absorbed by the electron sheet leading to a change in the 2-terminal
resistance which is recorded. Once the phonons have passed the 2DES they hit
the top face of the device where they undergo a variety of processes
including mode conversion, diffuse scattering and reflection. The top face
of the device is usually very close to the 2DES so that this occurs
virtually simultaneously with the encounter with the 2DES. The reflected
pulse travels back towards the rear face of the sample where it again
undergoes a variety of scattering/reflection processes. What is left of the
ballistic component of the pulse, having suffered two boundary reflections
heads back towards the 2DES. A clear response to the twice reflected phonons
can be seen in the experiments. This is an experimental verification of the
fact that the phonons of importance in these experiments really do propagate
ballistically. At longer times the ballistic component is essentially gone
and the time evolution of the 2 terminal voltage is simply due to the
general heating of the lattice by the energy dissipated in the heater; at
very long times the temperature control of the refrigeration system extracts
this energy, returning the system to its base temperature. The contacts that
feed current to the 2DES are deliberately placed far from it so that the
phonon pulses do not hit the contacts until long after they have passed the
2DES since the contacts will have a large response to the phonons.

\section{Electrons and Phonons in Heterostructures\label{Electrons}}

The electrons which accumulate at a GaAs/AlGaAs heterojunction at low
temperatures form a two-dimensional electron system\cite{AFS}. Each electron
has the same wave function for motion in the direction perpendicular to the
junction ($z$-direction) which we will denote by $f\left( z\right) $. The
acoustic phonons in GaAs are well described by the anisotropic Debye
approximation in which each phonon mode is labelled by a branch index $%
j=1,2,3$ (roughly speaking $1\times $ ``LA'' and $2\times $ ``TA'' modes)
and a 3d wave vector $\underline{Q}.$ A given mode has frequency 
\begin{equation}
\omega _{j}\left( \underline{Q}\right) =c_{j}\left( \underline{Q}/Q\right) Q
\end{equation}
and polarization $\xi ^{j}\left( \underline{Q}/Q\right) $ where $Q=\left| 
\underline{Q}\right| $ and $c_{j}\left( \underline{\widehat{n}}\right) $ is
a direction dependent speed of sound. In this paper we will neglect
focussing effects and use the isotropic approximation that replaces $%
c_{j}\left( \underline{Q}/Q\right) $ by a constant. This oversimplification
will be removed in subsequent numerical works. The main quantitative errors
introduced by this approximation are expected to be in the relative
magnitudes of the features associated with TA and LA\ phonons and their
detailed dependance on phonon angle of incidence.

There are two mechanisms by which electrons and phonons interact in GaAs\cite
{GandL}. Deformation potential coupling arises because the presence of a
long wavelength lattice distortion locally alters the band parameters,
particularly the energy of the conduction band edge. This is described by a
phenomenological expression for the potential energy experienced by an
electron 
\begin{equation}
V_{dp}\left( \underline{R}\right) =\Xi _{\mu \nu }S_{\mu \nu }\left( 
\underline{R}\right) \qquad \mu ,\nu =x,y,z
\end{equation}
where $\Xi _{\mu \nu }$ is the deformation potential tensor and $S_{\mu \nu
}\left( \underline{R}\right) $ is the lattice strain tensor. Another
coupling mechanism arises from the fact that, in materials whose lattice
structure does not possess a centre of symmetry, elastic distortions give
rise to local charge rearrangements which are characterized by an electric
polarization field 
\begin{equation}
P_{\mu }\left( \underline{R}\right) =\frac{1}{\kappa }h_{\mu \nu \lambda
}S_{\nu \lambda }\left( \underline{R}\right)
\end{equation}
where $\kappa $ is the relative permittivity of the material and $h_{\mu \nu
\lambda }$ is the piezomodulus tensor. This polarization field also gives
rise to a one-electron potential. In GaAs the deformation potential tensor
is actually a simple scalar 
\begin{equation}
\Xi _{\mu \nu }=\Xi _{0}\delta _{\mu \nu }
\end{equation}
with $\Xi _{0}\sim 7\mathop{\rm eV}$, while the piezomodulus tensor has the
form 
\begin{equation}
h_{\mu \nu \lambda }=\left\{ 
\begin{array}{c}
h_{14}\qquad \mu \neq \nu \neq \lambda \\ 
0\qquad \text{otherwise}
\end{array}
\right. \qquad .
\end{equation}
with $h_{14}\sim 0.14\mathop{\rm C}\mathop{\rm m}$.

In second quantized form, the combination of these mechanisms gives rise to
an electron-phonon coupling Hamiltonian of the form 
\begin{equation}
H=-\sum_{j,\underline{Q}}M_{j}\left( \underline{Q}\right) Z\left(
q_{z}\right) \widehat{\rho }_{-{\bf q}}\left( \widehat{a}_{j}^{\dagger
}\left( \underline{Q}\right) +\widehat{a}_{j}\left( -\underline{Q}\right)
\right)
\end{equation}
where $\widehat{a}_{j}^{\dagger }\left( \underline{Q}\right) $ creates a
phonon in the given mode, $\widehat{\rho }_{{\bf q}}$ is the Fourier
component of the (2d) electron density, $M_{j}\left( \underline{Q}\right) $
is the matrix element for phonon absorption/emission and $Z\left(
q_{z}\right) $ is a form factor for the finite thickness of the 2DES.
Details of the derivation of this form are presented in appendix \ref
{Couplings} along with the detailed expressions for $M_{j}\left( \underline{Q%
}\right) $. The simplified forms for the electron phonon couplings that
emerge from the isotropic Debye approximation are discussed in appendix \ref
{Debye}.

\section{The Fractional Quantum Hall Liquid}

The fractional quantum Hall effect was the first experimentally observed
manifestation of the incompressible quantum liquid state that occurs in 2d
electron systems subject to a strong magnetic field, $B$, such that the
Landau level filling factor $\nu =2\pi l_{c}^{2}\overline{\rho }$ ($%
\overline{\rho }$ is the area density of electrons and $l_{c}=\sqrt{\hbar /eB%
}$ is the magnetic length) is close to a low order, odd denominator fraction 
\cite{QHBook,FQHEBook,Stone,QHEBook2}. Laughlin\cite{Laughlin} proposed a
trial wave function for the ground state of this liquid at filling factors
of the form $\nu =1/\left( 2m+1\right) $ which clearly gave these fractions
an especially low energy. Laughlin also described the charged quasiparticle
excitations above this ground state as fractionally charged objects that
obey fractional statistics. These quasiparticles require a finite energy for
their creation, even in the thermodynamic limit, which accounts for the
incompressibility of the state. The energy gap for the creation of these
objects, $\Delta _{tr}$, determines the longitudinal conductivity of the
system which has an activated temperature dependence, $\sigma _{xx}\sim
\sigma _{0}e^{-\Delta _{tr}/T}$, in ideal systems. In real systems this form
is not observed at the lowest temperatures because of the Anderson
localization of the quasiparticles. Instead, conduction is supposed to occur
via variable range hopping.

Girvin, MacDonald and Platzman\cite{GMP} (GMP) turned their attention to the
possibility of low energy collective excitations of the Laughlin liquid.
They employed a modification of the technique used by Feynman\cite{Feynman}
to describe the collective modes of superfluid helium-4. They proposed
neutral excited states characterized by a 2d wave vector ${\bf q}$ of the
form 
\begin{equation}
\left| {\bf q}\right\rangle ={\cal N}_{{\bf q}}^{-1/2}\overline{\rho _{{\bf q%
}}}\left| \Psi _{L}\right\rangle
\end{equation}
where $\left| \Psi _{L}\right\rangle $ is Laughlin's ground state and ${\cal %
N}_{{\bf q}}^{-1/2}$ is a normalization factor. The operator $\overline{\rho
_{{\bf q}}}$ is the Fourier component of the electronic density, projected
onto the lowest Landau level (see \cite{GandJ} for details of this
projection). The expectation value of the energy in this state was expressed
exactly in terms of the lowest Landau level projected static structure
factor $\overline{s\left( {\bf q}\right) }.$ This quantity has an exact
relation to the ordinary static structure factor which is known from
Monte-Carlo hypernetted chain calculations on the Laughlin wave function.
The dispersion curve that arose from that analysis \cite{GMP} is presented
schematically in figure \ref{fig2}. This dispersion $\Delta \left( q\right) $
\ has a number of important features. The collective mode has a gap at all
wave vectors; unlike superfluid helium, there is no linearly dispersing,
gapless phonon regime. This is a manifestation of the incompressibility of
the state. The gap is a minimum, $\Delta ^{\ast }$, at a finite wave vector, 
$q^{\ast }$, which corresponds approximately to a wavelength equal to the
average inter-particle spacing. By analogy with superfluid Helium, the
excitations close to $q^{\ast }$ are referred to a magnetorotons.

The supposition that the trial states $\left| {\bf q}\right\rangle $ are
true eigenstates is equivalent to using the single mode approximation for
the dynamic structure factor\cite{Quantum liquids}: 
\begin{equation}
S\left( q,\omega \right) =\overline{s}\left( q\right) \delta \left( \omega
-\Delta \left( q\right) \right) \qquad \omega \ll \omega _{c}\qquad ,
\end{equation}
where $\omega _{c}=eB/m^{\ast }$ is the cyclotron frequency which determines
the Landau level spacing.

At large $q$ the collective mode is more properly thought of as an exciton
composed of a pair of oppositely charged Laughlin quasiparticles separated
by a distance $l_{c}^{2}q/\nu $\cite{GMP}. The energy of the exciton will be
essentially the mutual Coulomb energy of the charged constituents 
\begin{equation}
\Delta \left( q\right) \sim \Delta _{\infty }-\frac{\nu ^{3}e^{2}}{4\pi
\varepsilon _{0}\kappa l_{c}^{2}q}\qquad l_{c}q\rightarrow \infty
\end{equation}
where $\Delta _{\infty }$ is the energy needed to create an infinitely
separated pair of quasiparticles, i.e. $\Delta _{\infty }=\Delta _{tr}.$

Since, as we shall argue, the interaction between bulk acoustic phonons and
the Laughlin liquid is dominated by the creation of magnetorotons with wave
vectors close to $q^{\ast }$, we employ the standard roton form for the
energy in the vicinity of the minimum\cite{roton} 
\begin{eqnarray}
\Delta \left( q\right) &\sim &\Delta ^{\ast }+\frac{\left( q-q^{\ast
}\right) ^{2}}{2\mu }\qquad q\rightarrow q^{\ast } \\
\mu ^{-1} &=&\left( \frac{d^{2}\Delta \left( q\right) }{dq^{2}}\right)
_{q=q^{\ast }}\qquad .
\end{eqnarray}

From now on we will use units which are most natural for the system in
question by taking $\hbar =l_{c}={\cal E}_{c}=1$ where ${\cal E}_{c}$ is the
Coulomb energy scale for the system which in S.I.\ units has the form 
\begin{equation}
{\cal E}_{c}=\frac{e^{2}}{4\pi \varepsilon _{0}\kappa l_{c}}\qquad .
\end{equation}
The unit of velocity is then $v_{c}=l_{c}{\cal E}_{c}/\hbar $ which is
conveniently independent of magnetic field and density. In these reduced
units the average speeds of sound are $c_{1}\approx 0.0305$, $%
c_{2}=c_{3}\approx 0.0196$ while the magnetoroton minimum for $\nu =1/3$ is
(in the absence of any finite thickness, disorder and Landau level mixing
effects ) $\Delta ^{\ast }\approx 0.075$, $q^{\ast }\approx 1.3$. In a real
system the expectation is that $\Delta ^{\ast }$ will be reduced by all
these effects.

\section{Phonon Transport}

We can treat the propagation of a phonon pulse across a device
semiclassically. A formal solution of the phonon Boltzman equation is not
difficult but the absence of elastic scattering at the frequencies of
interest renders even this level of sophistication unnecessary. We assume
that the heater emits phonons isotropically in \underline{$Q$} -space. This
assumption is known to be incorrect in detail \cite{mismatch} but will not
lead to qualitative errors here. In fact the most obvious effect of the
anisotropic emission is the suppression of phonon emission at large angles
to the normal to the interface between heater and substrate, these phonons
would ``miss'' the 2DES system anyway (in a more careful treatment they
would miss by an even greater margin because of phonon focussing). The
anisotropy in the emissivity for LA phonons, $e_{1}\left( \theta ,\phi
\right) $, is generally weak. The anisotropy in the emissivity for the TA
modes, $e_{2}\left( \theta ,\phi \right) $ and $e_{3}\left( \theta ,\phi
\right) ,$ is stronger but the total TA emissivity, $e_{2}+e_{3}$, is only
weakly anisotropic so that, at least within the isotropic Debye
approximation used here, we do no real harm by neglecting it.

We describe the phonons in terms of wave packets whose spatial scale
determines a normalization volume $V$. When the phonons are created at the
heater they are characterized by a distribution function 
\begin{equation}
n_{j}^{\text{heater}}\left( \underline{Q}\right) =n_{B}\left( \omega
_{j}\left( \underline{Q}\right) /T_{\phi }\right) \vartheta \left(
q_{z}\right)
\end{equation}
where $n_{B}$ is the usual Bose-Einstein distribution function and the step
function simply accounts for the fact that only phonons propagating into the
sample are created (the sample is mounted in a vacuum). The determination of
which wave packets have reached the location of the 2DES at time $t$ is a
geometric problem. In principle the distribution of phonons in contact with
the 2DES is given by 
\begin{eqnarray}
n_{j}\left( \underline{Q};t\right) &=&\frac{n_{B}\left( \omega _{j}\left( 
\underline{Q}\right) /T_{\phi }\right) }{\left| v_{z}^{j}\left( \underline{Q}%
\right) \right| }\vartheta \left( t-\frac{L}{v_{z}^{j}\left( \underline{Q}%
\right) }\right) \vartheta \left( t_{p}+\frac{L}{v_{z}^{j}\left( \underline{Q%
}\right) }-t\right)  \nonumber \\
&&\times \int d^{2}{\bf r}\chi _{2}\left( {\bf r}\right) \chi _{h}\left( 
{\bf r}-\frac{{\bf v}^{j}\left( \underline{Q}\right) }{v_{z}^{j}\left( 
\underline{Q}\right) }L\right)
\end{eqnarray}
where $L$ is the thickness of the substrate, the group velocity associated
with mode $j,\underline{Q}$ is 
\begin{eqnarray*}
\underline{V^{j}}\left( \underline{Q}\right) &=&\left( {\bf v}^{j}\left( 
\underline{Q}\right) ,v_{z}^{j}\left( \underline{Q}\right) \right) \\
V_{\mu }^{j}\left( \underline{Q}\right) &=&\frac{\partial \omega _{j}\left( 
\underline{Q}\right) }{\partial Q_{\mu }}\qquad ,
\end{eqnarray*}
$t_{p}$ is the duration of the pulse and $\chi _{2}$ and $\chi _{h}$ are
characteristic functions for the 2DES and the heater (i.e. $\chi _{2}\left( 
{\bf r}\right) $ is unity if ${\bf r}$ is within the area of the 2DES and
zero otherwise etc.). The two step functions ensure that the front of the
pulse has reached the far side of the device but the back of the pulse has
not, while the integral ensures that the group velocity points from
somewhere on the heater to somewhere on the 2DES; the energy distribution is
still that of a black body because no inelastic processes have occurred. In
general $n_{j}\left( \underline{Q}\right) $ is a complicated object to
calculate because the group velocity is a non-trivial function of the wave
vector. In the isotropic approximation used here we have 
\[
V_{\mu }^{j}\left( \underline{Q}\right) =c_{j}\frac{Q_{\mu }}{Q} 
\]
so that 
\begin{eqnarray}
n_{j}\left( \underline{Q};t\right) &=&n_{B}\left( c_{j}Q/T_{\phi }\right) 
\frac{Q}{\left| c_{j}q_{z}\right| }\vartheta \left( t-\frac{QL}{c_{j}q_{z}}%
\right) \vartheta \left( t_{p}+\frac{QL}{c_{j}q_{z}}-t\right)  \nonumber \\
&&\times \int d^{2}{\bf r}\chi _{2}\left( {\bf r}\right) \chi _{h}\left( 
{\bf r}-\frac{{\bf q}}{q_{z}}L\right)  \nonumber \\
&=&n_{B}\left( c_{j}Q/T_{\phi }\right) \chi _{j}\left( {\bf q}%
,q_{z};t\right) \qquad
\end{eqnarray}
which defines the characteristic function $\chi _{j}\left( {\bf q}%
,q_{z};t\right) $ for the phonon pulse. In practice, what will matter for
the purposes of this paper are the maximum and minimum angles of incidence $%
\theta =\tan ^{-1}\left( \left| {\bf q}\right| /q_{z}\right) $ for which $%
\chi _{j}$ is non-zero at any instant.

\section{Phonon Absorption in the Fractional Quantum Hall Regime}

Let us now suppose that at some instant the phonons in contact with the 2DES
are characterized by a non-equilibrium distribution function $n_{j}\left( 
\underline{Q}\right) $ as described above and that the 2DES is in internal
equilibrium at some temperature $T_{e}\ll T_{\phi }$. In appendix \ref
{transferrate} it is shown that the rate at which energy is transferred from
the phonon pulse to the incompressible liquid is given by the golden rule
expression 
\begin{equation}
P=2\pi \overline{\rho }\Omega \sum_{j,\underline{Q}}\omega _{j}\left( 
\underline{Q}\right) \left| M_{j}\left( \underline{Q}\right) \right|
^{2}\left| Z\left( q_{z}\right) \right| ^{2}S\left( {\bf q},\omega
_{j}\left( \underline{Q}\right) \right) \left( n_{j}\left( \underline{Q}%
\right) -e^{-\omega _{j}\left( \underline{Q}\right) /T_{e}}\left(
1+n_{j}\left( \underline{Q}\right) \right) \right)
\end{equation}
where $M_{j}\left( \underline{Q}\right) $ is the electron-phonon matrix
element discussed above and $S\left( {\bf q},\omega \right) $ is the dynamic
structure factor of the electron liquid. In appendix \ref{Structure} it is
shown that in the regime $T_{e}\ll \Delta ^{\ast }$, the leading asymptotics
of $P$ can be obtained by using the zero temperature form for the structure
factor for which we will use the single mode approximation \cite{GMP}. Hence
we find 
\begin{equation}
P=2\pi \overline{\rho }\Omega \sum_{j,\underline{Q}}\omega _{j}\left( 
\underline{Q}\right) \left| M_{j}\left( \underline{Q}\right) \right|
^{2}\left| Z\left( q_{z}\right) \right| ^{2}\overline{s}\left( {\bf q}%
\right) \left( n_{j}\left( \underline{Q}\right) -e^{-\omega _{j}\left( 
\underline{Q}\right) /T_{e}}\left( 1+n_{j}\left( \underline{Q}\right)
\right) \right) \delta \left( \omega _{j}\left( \underline{Q}\right) -\Delta
\left( q\right) \right)
\end{equation}
which has the straightforward interpretation that the processes involved are
the direct conversion between phonons and magnetorotons with the matrix
element $\left| M_{j}\left( \underline{Q}\right) \right| ^{2}\left| Z\left(
q_{z}\right) \right| ^{2}\overline{s}\left( {\bf q}\right) $. The two terms
within the bracket involving the phonon occupation numbers represent phonon
absorption and emission respectively. We are concentrating on the regime $%
T_{e},T_{\phi }\ll \Delta ^{\ast }$ which ensures that $n_{j}\left( 
\underline{Q}\right) \ll 1$ for all modes that have sufficient energy to
couple to the collective mode so that stimulated emission is always
negligible.

The phonon-magnetoroton conversion process is subject to the conservation of
in-plane wave vector and energy, so that when a phonon with wave vector $%
\underline{Q}=\left( {\bf q},q_{z}\right) $ is absorbed, a magnetoroton of
wave vector ${\bf q}$ is created only when $\omega \left( \underline{Q}%
\right) =\Delta \left( q\right) $. It is often very useful to visualize such
processes in terms of the crossing of dispersion curves. Here there is the
subtlety that the magnetoroton energy depends on the 2d wave vector $q$
while the phonon energy depends on the 3d wave vector $\underline{Q}$. We
can however create such a picture by concentrating on phonons with a given
angle of incidence $\theta $ (see the inset to figure \ref{schematic}). For
these phonons the energy depends on the in-plane component of the wave
vector via 
\begin{equation}
\omega =c_{j}Q=\frac{c_{j}q}{\sin \theta }\qquad .
\end{equation}
In figure \ref{fig2} the dispersion relation for the magnetorotons (in the
ideal case for the case $\nu =1/3$) is plotted along with the dispersion
curves for LA\ and TA\ phonons incident at $\theta =30^{\circ }$. In this
case the LA\ phonon with the quoted angle of incidence and $q=1.25$ or $%
q=1.65$ couple to the electron liquid but no TA\ phonons at this angle can.
(The dispersion curves will eventually cross because the $\Delta \left(
q\right) $ curve flattens out at larger $q,$ but this crossing will happen
at such large $q$ that the projected static structure factor, $\overline{s}%
\left( q\right) $, which decays as $e^{-q^{2}}$ for $ql_{c}\gg 1$, will be
negligibly small.) The actual location of the crossing point, of course
depends on the energy at the magnetoroton minimum. In real systems there are
three effects which all act to reduce the size of the magnetoroton gap:
finite thickness modifications to the effective inter-electron interaction;
Landau level mixing and disorder \cite{disorder}.

\section{Asymptotic Expressions for The Energy Transfer Rate}

Let us concentrate on the regime $T_{\phi },T_{e}\ll \Delta ^{\ast }$. In
this case we can neglect stimulated phonon emission and replace the
Bose-Einstein factor $n_{B}\left( \omega /T_{\phi }\right) $ by the
exponential $e^{-\omega /T_{\phi }}$ for all phonons that can couple to the
electron liquid. Hence we obtain, converting the sum over wave vectors into
an integral, 
\begin{eqnarray}
P &\sim &\overline{\rho }\Omega \frac{V}{4\pi ^{2}}\sum_{j}\int d^{3}%
\underline{Q}\omega _{j}\left( \underline{Q}\right) \left| M_{j}\left( 
\underline{Q}\right) \right| ^{2}\left| Z\left( q_{z}\right) \right| ^{2}\,%
\overline{s}\left( q\right)  \nonumber \\
&&\times \left( e^{-\omega _{j}\left( \underline{Q}\right) /T_{\phi }}\chi
_{j}\left( {\bf q},q_{z}\right) -e^{-\omega _{j}\left( \underline{Q}\right)
/T_{e}}\right) \delta \left( \omega _{j}\left( \underline{Q}\right) -\Delta
\left( q\right) \right) \qquad .
\end{eqnarray}
At this point we will make a drastic simplification which will surely lead
to quantitative error but will allow swift analytic progress and qualitative
understanding: we will make the isotropic Debye approximation and set $%
\omega _{j}\left( \underline{Q}\right) =c_{j}Q$. As can be seen in appendix 
\ref{Debye} this greatly simplifies the form of $\left| M_{j}\left( 
\underline{Q}\right) \right| ^{2}$ leading to 
\[
P=P_{LA}^{+}-P_{LA}^{-}+P_{TA}^{+}-P_{TA}^{-} 
\]
\begin{eqnarray}
P_{s}^{\pm } &\sim &\overline{\rho }\Omega \frac{Vc_{s}}{4\pi ^{2}}%
\int_{0}^{\infty }dQQ^{3}e^{-c_{s}Q/T_{\pm }}\int_{0}^{\pi }d\theta \sin
\theta \left| Z\left( Q\cos \theta \right) \right| ^{2}\overline{s}\left(
Q\sin \theta \right) \delta \left( c_{s}Q-\Delta \left( Q\sin \theta \right)
\right)  \nonumber \\
&&\times \int_{0}^{2\pi }d\phi \Gamma _{s}\left( Q,\sin \theta ,\phi \right)
\chi _{j}^{\pm }\left( Q,\sin \theta ,\phi \right) \qquad s=LA,TA.
\end{eqnarray}
where $\chi _{j}^{+}=\chi _{j}$, $\chi _{j}^{-}=1$, $T_{+}=T_{\phi }$ and $%
T_{-}=T_{e}$. The $\phi $ integral averages the electron phonon coupling
over available in-plane directions, hence we define 
\[
\widetilde{\Gamma _{s}^{\pm }}\left( Q,\sin \theta \right) =\frac{1}{2\pi }%
\int_{0}^{2\pi }d\phi \Gamma _{s}\left( Q,\sin \theta ,\phi \right) \chi
_{j}^{\pm }\left( Q,\sin \theta ,\phi \right) \qquad . 
\]
leaving us with the simplified expression 
\begin{equation}
P_{s}^{\pm }=\overline{\rho }\Omega \frac{Vc_{s}}{2\pi }\int_{0}^{\infty
}dQQ^{3}e^{-c_{s}Q/T_{\pm }}\int_{0}^{\pi }d\theta \sin \theta \widetilde{%
\Gamma _{s}^{\pm }}\left( Q,\sin \theta \right) \left| Z\left( Q\cos \theta
\right) \right| ^{2}\overline{s}\left( Q\sin \theta \right) \delta \left(
c_{s}Q-\Delta \left( Q\sin \theta \right) \right)
\end{equation}
which can be recast into the form (see appendix \ref{integral}) 
\begin{eqnarray}
P_{s}^{\pm } &\sim &\frac{\overline{\rho }\Omega V}{2\pi c_{s}^{2}}%
\int_{0}^{\infty }dqq\int_{c_{s}q}^{\infty }d\omega \frac{\omega ^{2}}{%
q_{s}^{z}\left( \omega ,q\right) }\widetilde{\Gamma _{s}^{\pm }}\left( \frac{%
\omega }{c_{s}},\frac{c_{s}q}{\omega }\right) \left| Z\left( q_{s}^{z}\left(
\omega ,q\right) \right) \right| ^{2}  \nonumber \\
&&\times \overline{s}\left( q\right) e^{-\omega /T_{\pm }}\delta \left(
\omega -\Delta \left( q\right) \right)
\end{eqnarray}
where we have used the convenient shorthand notation 
\begin{equation}
q_{s}^{z}\left( \omega ,q\right) =\frac{\sqrt{\omega ^{2}-c_{s}^{2}q^{2}}}{%
c_{s}}
\end{equation}
Carrying out the trivial $\omega $ integration then gives 
\begin{eqnarray}
P_{s}^{\pm } &\sim &\int F_{s}^{\pm }\left( q\right) \overline{s}\left(
q\right) e^{-\Delta \left( q\right) /T_{\pm }}dq\qquad T_{+}=T_{\phi }\qquad
T_{-}=T_{e} \\
F_{s}^{\pm }\left( q\right) &=&\frac{V\overline{\rho }\Omega }{2\pi }\left( 
\frac{\Delta \left( q\right) }{c_{s}}\right) ^{2}\widetilde{\Gamma _{s}^{\pm
}}\left( \frac{\Delta \left( q\right) }{c_{s}},\frac{c_{s}q}{\Delta \left(
q\right) }\right) \left| Z\left( q_{s}^{z}\left( \Delta \left( q\right)
,q\right) \right) \right| ^{2}q\frac{\vartheta \left( \Delta \left( q\right)
-c_{j}q\right) }{q_{s}^{z}\left( \omega ,q\right) }
\end{eqnarray}
which is still a rather formidable integral. However we can make use of the
fact that we are in the regime $T_{e},T_{\phi }\ll \Delta ^{\ast }$ and use
the method of steepest descents\cite{BandO} to establish the leading
behaviour of the integral. Naively, we would expect the integrals to be
dominated by the minimum in the magnetoroton dispersion curve at $q^{\ast }$
but this is only true for the absorption integral provided $\widetilde{%
\Gamma _{s}^{\pm }}\left( \frac{\Delta \left( q^{\ast }\right) }{c_{s}},%
\frac{c_{s}q^{\ast }}{\Delta \left( q^{\ast }\right) }\right) \neq 0$, \ in
other words that there are phonons in the pulse in contact with the
electrons that can couple at the magnetoroton minimum. If we characterize
the pulse by maximum and minimum angles of incidence $\theta _{\max }$ and $%
\theta _{\min }$ then we can distinguish three cases.

\begin{enumerate}
\item  In the case that $c_{j}q^{\ast }/\sin \theta _{\max }<\Delta \left(
q^{\ast }\right) <c_{j}q^{\ast }/\sin \theta _{\min }$ then there are
phonons in the pulse that can couple to the magnetoroton minimum and the
absorption integral will be dominated by the point $q^{\ast }$: we refer to
this as type I absorption.

\item  In the case that $\Delta \left( q^{\ast }\right) <c_{j}q^{\ast }/\sin
\theta _{\max }<c_{j}q^{\ast }/\sin \theta _{\min }$ there are no phonons in
the pulse that can couple to the magnetoroton minimum and the absorption
integral is dominated by coupling to the collective mode with the lowest
energy which can couple to phonons in the pulse, these have wave vector
given by the lowest energy solution of 
\begin{equation}
\Delta \left( q\right) \sin \theta _{\max }=c_{j}q\qquad ;  \label{dominant}
\end{equation}
this solution will always have $q<q^{\ast }$: we refer to this as type II
absorption.

\item  In the case that $c_{j}q^{\ast }/\sin \theta _{\max }<c_{j}q^{\ast
}/\sin \theta _{\min }<\Delta \left( q^{\ast }\right) $ there are simply no
phonons in the pulse that can couple to the electronic system and there is
no absorption (type III).
\end{enumerate}

For the emission integral the same cases apply except that $\theta _{\max
}=\pi /2$ and $\theta _{\min }=0$ (so that case 3 does not arise). This of
course ignores the possibility that for some directions $\left( \theta ,\phi
\right) $ the characteristic function $\chi $ for the pulse may be non-zero
but the coupling to the electrons may vanish because of a zero in the $%
A\left( \theta \right) $ or $B\left( \phi \right) $ coefficients defined in
appendix \ref{Debye}.

We thus have two integrals to estimate 
\begin{equation}
\int_{0}^{\infty }F_{j}^{\pm }\left( q\right) e^{-\Delta \left( q\right)
/T}dq
\end{equation}
and 
\begin{equation}
\int_{0}^{q_{0}}F_{j}^{\pm }\left( q\right) e^{-\Delta \left( q\right)
/T}dq\qquad q_{0}<q^{\ast }\qquad .
\end{equation}
In the first case we expand around the minimum and find 
\begin{eqnarray}
P_{j}^{+} &\sim &\int F_{j}^{\pm }\left( q\right) e^{-\Delta \left( q\right)
/T_{\pm }}dq\sim F_{j}^{\pm }\left( q^{\ast }\right) e^{-\Delta ^{\ast
}/T_{\pm }}\int_{-\infty }^{\infty }e^{-x^{2}/2\mu T_{\pm }}dx  \nonumber \\
&=&\sqrt{2\pi \mu T_{\pm }}F_{j}^{\pm }\left( q^{\ast }\right) e^{-\Delta
^{\ast }/T_{\pm }}\qquad T_{\pm }\rightarrow 0.
\end{eqnarray}
In the second case the integral is dominated by the upper end-point and we
must expand around here 
\begin{equation}
\Delta \left( q\right) \sim \Delta \left( q_{0}\right) +v_{q}\left(
q-q_{0}\right)
\end{equation}
where, because $q_{0}<q^{\ast }$, we will have 
\begin{equation}
v_{q}=\left( \frac{d\Delta \left( q\right) }{dq}\right) _{q=q_{0}}<0\qquad .
\end{equation}
Hence 
\begin{eqnarray}
P_{j}^{+} &\sim &\int_{0}^{q_{0}}F_{j}^{\pm }\left( q\right) e^{-\Delta
\left( q\right) /T_{\pm }}dq\sim F_{j}^{\pm }\left( q_{0}\right) e^{-\Delta
\left( q_{0}\right) /T_{\pm }}\int_{0}^{\infty }e^{-\left| v_{q_{0}}\right|
x/T_{\pm }}dx  \nonumber \\
&=&\frac{T_{\pm }}{\left| v_{q_{0}}\right| }F_{j}^{\pm }\left( q_{0}\right)
e^{-\Delta \left( q_{0}\right) /T_{\pm }}\qquad T_{\pm }\rightarrow 0\qquad .
\end{eqnarray}

From the above analysis we can see that, at least in principal, phonon
absorption experiments have the potential to reveal detailed spectroscopic
information about the magnetoroton dispersion curve, at least in the
vicinity of the minimum. By choosing the geometry of the experiment such
that one is the situation where the phonon band delimited by $\theta _{\max
} $ and $\theta _{\min }$ does not include the magnetoroton minimum, the
rate of energy transfer should be dominated by phonons with the maximum
angle of incidence.

In order to illustrate these ideas let us consider the sample geometry shown
in figure \ref{schematic} and estimate the response to the angled heater H2,
taking the maximum and minimum angles of incidence to be $\theta _{\min
}=\pi /5$ and $\theta _{\max }=2\pi /7.$ Suppose that the magnetoroton gap
is somewhat smaller than its ideal value, say $\Delta ^{\ast }=0.05{\cal E}%
_{c}$. Figure \ref{fig3a} shows the extreme LA lines on top of the
magnetoroton dispersion curve while figure \ref{fig3b} shows the extreme TA
lines. From these we can see that there will be no TA absorption (type III)
while the LA\ absorption is (just) type II with $q_{0}$ very close to $%
q^{\ast }$. Of course the details of which type of process dominates in a
given geometry will depend on the actual phonon dispersion relations
including the effects of anisotropy.

\section{Electron Heating}

Let us now consider in more detail the experimental situation. In the
experiments the quantity that is directly measured is the two-terminal
resistance of the device as a function of time. We appeal to the wide
separation between the timescales for the electron system ($\sim 10^{-14}s$)
and that of the experiment ($\sim 10^{-8}s$) and the relatively weak
coupling to the phonons to assert that on experimentally resolvable
timescales the electronic system is always in local equilibrium at some well
defined temperature $T_{e}$. We suppose then, that the effect of the phonons
is to heat up the electron gas and all that we observe is a consequence of
the time variation of the electron temperature $T_{e}\left( t\right) $. The
resistance of the sample as a function of temperature can be obtained in a
straightforward transport experiment in which the equilibrium temperature of
the whole device is varied while the current is kept constant. Hence the
results of the phonon experiments can be converted directly into a plot of $%
T_{e}\left( t\right) $.

The time dependence of the electron temperature is given by 
\begin{equation}
\frac{dT_{e}\left( t\right) }{dt}=\frac{P\left( T_{\phi },T_{e}\left(
t\right) \right) }{C\left( T_{e}\left( t\right) \right) }
\end{equation}
where $P$ is the rate of energy transfer to the 2DES and $C\left(
T_{e}\right) $ is the heat capacity of the electronic system at temperature $%
T_{e}$. We will focus on the early time behaviour when $T_{e}\ll T_{\phi
}\ll \Delta ^{\ast }$ and suppose that the geometry of the heater and active
device are such that the absorption is type I. If we suppose that we have a
uniform system in the $\nu =1/m$ ($m$ odd) state then we have 
\[
C\left( T_{e}\right) =C_{0}e^{-\Delta ^{\ast }/T_{e}} 
\]
for the electron gas itself. However the electron gas will be in thermal
contact with metallic contacts with heat capacity $A_{\text{contact}}T_{e}$
so that we assume 
\[
\frac{dT_{e}\left( t\right) }{dt}\sim \frac{\widetilde{F}\sqrt{T_{\phi }}%
e^{-\Delta ^{\ast }/T_{\phi }}}{C_{0}e^{-\Delta ^{\ast }/T_{e}\left(
t\right) }+A_{\text{contact}}T_{e}\left( t\right) }\qquad . 
\]
At early times the denominator will be dominated by the linear term so that
we expect 
\begin{eqnarray*}
T_{e}\left( t\right) &\sim &\sqrt{T_{0}^{2}+\left( 2\frac{\widetilde{F}}{A}%
\sqrt{T_{\phi }}e^{-\Delta ^{\ast }/T_{\phi }}\right) t} \\
&\sim &T_{0}\left( 1+\left( \frac{\widetilde{F}}{A}\sqrt{T_{\phi }}%
e^{-\Delta ^{\ast }/T_{\phi }}\right) t\right) \qquad t\rightarrow 0\qquad .
\end{eqnarray*}

In a real high mobility device the 2DES is subject to potential fluctuations
which are slow on the scale of the magnetic length $l_{c}$ so that their
effect can be described within the Thomas Fermi approximation \cite
{Efros,Chklovskii,Chalker,Chklovskii/Lee}. This predicts that, in the
quantum Hall regime with $\nu =1/2m+1$, the bulk of the sample will consist
of a percolating `lake' of incompressible fluid. Embedded within the lake
will be isolated `islands' of compressible fluid (perhaps themselves
containing fully depleted regions). Similarly at the boundary of the active
area there will be strips of compressible fluid between the incompressible
lake and the fully depleted region outside the device. Let $\eta $ be the
fraction of the area occupied by the electrons which is covered by the
percolating incompressible liquid then we write 
\begin{eqnarray}
P\left( T_{\phi },T_{e}\left( t\right) \right) &=&\eta P_{i}\left( T_{\phi
},T_{e}\left( t\right) \right) +\left( 1-\eta \right) P_{c}\left( T_{\phi
},T_{e}\left( t\right) \right) \\
C\left( T_{e}\left( t\right) \right) &=&\eta C_{i}\left( T_{e}\left(
t\right) \right) +\left( 1-\eta \right) C_{c}\left( T_{e}\left( t\right)
\right)
\end{eqnarray}
where $P_{i}$ and $C_{i}$ are the energy transfer rate and the heat capacity
of the incompressible fluid and $P_{c}$ and $C_{c}$ the corresponding
quantities for the compressible regions. In the samples used in the
experiments $\eta $ may not be so small because the active 2DES is patterned
into a meander line to increase its overall resistance and hence its
sensitivity as a bolometer. In the previous sections we have developed a
theory for $P_{i}$. Its counterpart in the compressible regions\cite
{Benedict} will have the form 
\begin{equation}
P_{c}\left( T_{\phi },T_{e}\left( t\right) \right) \sim U_{0}\left( T_{\phi
}^{n}-\left( T_{e}\left( t\right) \right) ^{n}\right)
\end{equation}
where the power $n$ depends on the dominant coupling process and on the
nature of the correlations in the compressible state. The heat capacity of
the compressible regions will also be metallic with 
\begin{equation}
C_{c}\left( T_{e}\right) =A_{c}T_{e}\qquad .
\end{equation}
Hence we have that 
\begin{equation}
\frac{dT_{e}\left( t\right) }{dt}=\frac{\eta F\sqrt{T_{\phi }}e^{-\Delta
^{\ast }/T_{\phi }}+\left( 1-\eta \right) U_{0}\left( T_{\phi }^{n}-\left(
T_{e}\left( t\right) \right) ^{n}\right) }{\eta C_{0}e^{-\Delta ^{\ast
}/T_{e}}+\left[ \left( 1-\eta \right) A_{c}+A_{\text{contact}}\right] T_{e}}%
\qquad .
\end{equation}
Thus we see that the presence of compressible regions can greatly complicate
the interpretation of results.

\section{Electron Equilibration}

All of the above is based on the supposition that there is a wide separation
between the timescale for experimental measurements of the resistance and
the internal (energy) relaxation time for the 2DES. Normally one supposes
that electronic relaxation times are of the order of $10^{-15}s$ while the
instrumental resolution of current phonon experiments is of the order of a
nanosecond. Our confidence in the belief that the electronic systems is
always in local equilibrium at temperature $T_{e}$ is based on recent
systematic experiments which show that the results obtained in the phonon
absorption experiments do not have any significant dependence on the
duration of the phonon pulse $t_{p}$\cite{newprl}. We will, however, at
least consider the equilibration processes involved.

Equilibration within the compressible regions will be dominated by
quasiparticle scattering and therefore rapid except at the very lowest
temperatures. Equilibration within the incompressible regions is more
problematic. One could imagine a process whereby the creation of
magnetorotons by the absorption of phonons from the high energy tail of the
Bose distribution leads directly to the formation of unbound charged
quasiparticles. This process would involve a magnetoroton being scattered
into a higher energy state by the absorption of a low energy phonon from the
bulk of the Bose distribution and that this gives rise to a steady
``heating'' of a magnetoroton until its size exceeds some screening length
at which point it dissociates into a pair of oppositely charged
quasiparticles. A simple estimate\cite{Wurtzburg} for the rate of this kind
of process was made which simply assumed a magnetoroton ``density of
states'' with the result that the growth in the number of quasiparticles was
sub-linear ($\sim \sqrt{t}$). A more careful analysis of this process
including phase space restrictions on magnetoroton-phonon scattering\cite
{BHB}, however, shows that this process will not heat the magnetorotons to
sizes in excess of $\sim 3l_{c}$ and hence will not lead to any free
quasiparticles. This process could of course still dominate in ``dirty''
situations in which conservation of magnetoroton momentum is not necessary
due to the effects of short range correlated disorder. The dominant
mechanism for equilibration is therefore likely to involve collisions
between magnetorotons. This rate should be proportional to $e^{-2\Delta
^{\ast }/T_{e}}$ and so could become rather slow at low electron
temperatures. Equilibration between the percolating sea of Laughlin liquid
and the compressible regions (including the contacts) is presumably by
conversion processes in which a ballistically propagating magnetoroton
encounters a compressible region and is converted into a particle-hole
excitation.

\section{Summary}

In this paper we have developed a basic framework for the interpretation of
phonon absorption experiments in the fractional quantum Hall regime. Our
basic picture is the following. Phonons with a black-body distribution of
energies are injected into the GaAs substrate of a device containing a 2DES
at a GaAs/AlGaAs heterojunction. The phonons propagate ballistically with
spatiotemporal separation of the ``LA'' and ``TA'' phonons. Those phonons
that have the appropriate group velocities eventually come into contact with
the 2DES. The transfer of energy from the part of the phonon pulse that
meets the electron gas is controlled by the process in which phonons are
destroyed and magnetorotons created. The energy that is transferred to the
electronic system via this bottleneck rapidly equilibrates, with metallic
contacts (and possibly islands of compressible fluid) acting as thermal
reservoirs, leaving the electronic system at a raised temperature with a
consequent increase in the two-terminal resistance. We predict that the rate
at which the electronic system absorbs energy from the phonon pulse will
have a dependence on the heater temperature of the form $\sqrt{T_{\phi }}%
e^{-\Delta ^{\ast }/T_{\phi }}$ where $\Delta ^{\ast }$ is the gap at the
magnetoroton minimum, provided there are phonons in the pulse that can
couple at the minimum (type I absorption). If this is not the case then the
energy absorption rate will be dominated by the lowest energy magnetorotons
to which the phonons can couple and will have the form $T_{\phi }e^{-\Delta
\left( q_{0}\right) /T_{\phi }}$ where $q_{0}$ is the lowest energy solution
of equation \ref{dominant}. Hence, by judiciously positioning heaters on the
bottom face of a device so that a range of angles can be sampled, it should
be possible to map the dispersion relation $\Delta (q)$ for $q\lesssim
q^{\ast }$ at appropriate discrete $q$ values. It has also been argued that
the presence of compressible regions within the active area of the 2DES will
introduce additional phonon absorption with a power law dependence on the
heater temperature.

All of the detailed calculations in the preceding sections have been carried
out in the asymptotic regime $T_{e},T_{\phi }\ll \Delta ^{\ast }$ and have
made isotropic approximations for the propagation of the phonons. Numerical
calculations are in progress to rectify these deficiencies and these will
appear as a subsequent publication Another shortcoming of the present work
is that we have based everything on the Single-Mode Approximation of Girvin,
MacDonald and Platzman which is only reliable for fractional states with
filling factors $\nu =1/3$, $1/5$ etc. Work is in progress to investigate
the use of other frameworks for the collective modes of fractional quantum
Hall systems based on composite fermion ideas\cite{Simon}.\medskip 

{\em This work was supported by the EPSRC(U.K.) under grant number
GR/K41168. We are grateful to the following for many useful discussions: M.
Brownlie, A.J. Kent, U. Zeitler, L.J. Challis, W. Dietsche, V.I. Fal'ko, D.
Lehmann, A Devitt, J.E.Digby and R.H. Eyles. We are grateful to the Referee
for pointing out a false assumption in an earlier version of this paper.}

\appendix

\section{Electron-Phonon Coupling}

\label{Couplings}The two mechanisms by which the electrons and phonons
couple lead to similar forms of Hamiltonian, fortunately they appear $\pi /2$
out of phase so that they do not interfere with one another quantum
mechanically. We will consider first the deformation potential coupling and
then the piezoelectric coupling, however a number of preliminary quantities
are used in both. The second quantized Hamiltonian for electrons subject to
a one-body potential $V\left( \underline{R}\right) $ has the form 
\begin{equation}
H=\int V\left( \underline{R}\right) \widehat{\rho }\left( \underline{R}%
\right) d^{3}\underline{R}
\end{equation}
where $\widehat{\rho }\left( \underline{R}\right) $ is the electron density.
In order that we only have to deal with two-dimensional quantities when
referring to the 2DES, it is useful to take the expectation value with
respect to the vertical ($z$) part of the electron wave function so that 
\begin{eqnarray}
H &=&\int \widehat{\rho }\left( {\bf r}\right) \left\{ \int V\left( {\bf r}%
,z\right) \left| f\left( z\right) \right| ^{2}dz\right\} d^{2}{\bf r} 
\nonumber \\
&=&\int \widehat{\rho }\left( {\bf r}\right) U\left( {\bf r}\right) d^{2}%
{\bf r}
\end{eqnarray}
where $\widehat{\rho }\left( {\bf r}\right) $ is the area density of
electrons in the lowest vertical sub-band. It is convenient to work in the
Fourier representation 
\begin{equation}
\widehat{\rho }\left( {\bf r}\right) =\frac{1}{\Omega }\sum_{{\bf q}}%
\widehat{\rho }_{{\bf q}}e^{i{\bf q\cdot r}}\qquad \rho _{{\bf q}}=\int \rho
\left( {\bf r}\right) e^{-i{\bf q\cdot r}}d^{2}{\bf r}
\end{equation}
where $\Omega $ is the area of the 2DES. In this connection we will often
require the form factor 
\begin{equation}
Z\left( q_{z}\right) =\int \left| f\left( z\right) \right|
^{2}e^{iq_{z}z}dz\qquad .
\end{equation}
It is common to use the Fang-Howard form 
\begin{equation}
f\left( z\right) =\frac{1}{\sqrt{2a^{3}}}ze^{-z/2a}\vartheta \left( z\right)
\end{equation}
for a heterojunction (with the AlGaAs on the $z<0$ side) where $a$ is a
measure of the ``thickness'' of the 2DES in which case 
\begin{equation}
Z\left( q_{z}\right) =\left( 1+iaq_{z}\right) ^{-3}\qquad .
\end{equation}

The elastic displacement at position $\underline{R}$ is written in terms of
phonon operators as 
\begin{eqnarray}
u_{\mu }\left( \underline{R}\right) &=&\sum_{j,\underline{Q}}u_{\mu
}^{j}\left( \underline{Q}\right) e^{i\underline{Q}\cdot \underline{R}} \\
u_{\mu }^{j}\left( \underline{Q}\right) &=&\sqrt{\frac{\hbar }{2N_{0}V\omega
_{j}\left( \underline{Q}\right) }}\xi _{\mu }^{j}\left( \underline{Q}\right)
\left( \widehat{a}_{j}^{\dagger }\left( \underline{Q}\right) +\widehat{a}%
_{j}\left( -\underline{Q}\right) \right)
\end{eqnarray}
where $N_{0}$ is the density of the material ($5.14kgm^{-3}$ for GaAs) and $%
V $ is a normalization volume. The strain field is then given by 
\begin{eqnarray}
{\cal S}_{\mu \nu }\left( \underline{R}\right) &=&\frac{1}{2}\left( \frac{%
\partial u_{\mu }\left( \underline{R}\right) }{\partial R_{\nu }}+\frac{%
\partial u_{\nu }\left( \underline{R}\right) }{\partial R_{\mu }}\right) \\
&=&\sum_{j,\underline{Q}}{\cal S}_{\mu \nu }^{j}\left( \underline{Q}\right)
e^{i\underline{Q}\cdot \underline{R}} \\
{\cal S}_{\mu \nu }^{j}\left( \underline{Q}\right) &=&\frac{i}{2}\sqrt{\frac{%
\hbar }{2N_{0}V\omega _{j}\left( \underline{Q}\right) }}\left( \xi _{\mu
}^{j}\left( \underline{Q}\right) Q_{\nu }+Q_{\mu }\xi _{\nu }^{j}\left( 
\underline{Q}\right) \right) \left( \widehat{a}_{j}^{\dagger }\left( 
\underline{Q}\right) +\widehat{a}_{j}\left( -\underline{Q}\right) \right)
\qquad .
\end{eqnarray}

\subsection{Deformation Potential Coupling}

The deformation potential due to a lattice strain ${\cal S}_{\mu \nu }$ is 
\begin{equation}
V_{dp}\left( \underline{R}\right) =\Xi _{\mu \nu }{\cal S}_{\mu \nu }\left( 
\underline{R}\right)
\end{equation}
so that, in second quantization, we have 
\begin{eqnarray}
\widehat{H}_{dp} &=&\int U_{dp}\left( {\bf r}\right) \widehat{\rho }\left( 
{\bf r}\right) d^{2}{\bf r} \\
&=&\Xi _{\mu \nu }\int \widehat{\rho }\left( {\bf r}\right) \int \left|
f\left( z\right) \right| ^{2}\widehat{{\cal S}}_{\mu \nu }\left( {\bf r}%
,z\right) dzd^{2}{\bf r} \\
&=&i\sum_{j,\underline{Q}}\Xi _{j}\left( \underline{Q}\right) Z\left(
q_{z}\right) \widehat{\rho }_{-{\bf q}}\left( \widehat{a}_{j}^{\dagger
}\left( \underline{Q}\right) +\widehat{a}_{j}\left( -\underline{Q}\right)
\right)
\end{eqnarray}
where 
\begin{equation}
\Xi _{j}\left( \underline{Q}\right) =\frac{1}{2}\Xi _{\mu \nu }\sqrt{\frac{%
\hbar }{2N_{0}V\omega _{j}\left( \underline{Q}\right) }}\left( \xi _{\mu
}^{j}\left( \underline{Q}\right) Q_{\nu }+Q_{\mu }\xi _{\nu }^{j}\left( 
\underline{Q}\right) \right)
\end{equation}
which, for GaAs, reduces to 
\begin{equation}
\Xi _{j}\left( \underline{Q}\right) =\Xi _{0}\sqrt{\frac{\hbar }{%
2N_{0}V\omega _{j}\left( \underline{Q}\right) }}\left( \underline{\xi }%
^{j}\left( \underline{Q}\right) \cdot \underline{Q}\right) \qquad .
\end{equation}
The conventional value for the deformation potential parameter is $\Xi
_{0}\approx 7eV$ but the variance between different measurements is quite
large.

\subsection{Piezoelectric Coupling}

The polarization due to a lattice strain ${\cal S}_{\mu \nu }$ is 
\begin{equation}
P_{\mu }\left( \underline{R}\right) =\frac{1}{\kappa }h_{\mu \nu \lambda }%
{\cal S}_{\nu \lambda }\left( \underline{R}\right) \qquad .
\end{equation}
The electrostatic potential associated with this polarization field is $\phi
\left( \underline{R}\right) $ which corresponds to an electric field $%
\underline{E}=-\nabla \phi $. These are related by the Maxwell equation 
\begin{equation}
\nabla \cdot \underline{D}=\nabla \cdot \left( \varepsilon _{0}\underline{E}+%
\underline{P}\right) =0
\end{equation}
so that 
\begin{equation}
\varepsilon _{0}\nabla ^{2}\phi =\nabla \cdot \underline{P}\qquad .
\end{equation}
Now 
\begin{equation}
\nabla \cdot \underline{P}\left( \underline{R}\right) =\frac{i}{\kappa }%
h_{\mu \nu \lambda }\sum_{j,\underline{Q}}Q_{\mu }{\cal S}_{\nu \lambda
}^{j}\left( \underline{Q}\right) e^{i\underline{Q}\cdot \underline{R}}
\end{equation}
so that 
\begin{equation}
\phi \left( \underline{R}\right) =\sum_{\underline{Q}}\phi \left( \underline{%
Q}\right) e^{i\underline{Q}\cdot \underline{R}}\qquad
\end{equation}
where 
\begin{equation}
\phi \left( \underline{Q}\right) =-ih_{\mu \nu \lambda }\frac{Q_{\mu
}\sum_{j}{\cal S}_{\nu \lambda }^{j}\left( \underline{Q}\right) }{\kappa
\varepsilon _{0}Q^{2}}\qquad .
\end{equation}
The electron phonon Hamiltonian is then 
\begin{eqnarray}
\widehat{H}_{pa} &=&-e\int \widehat{\phi }\left( \underline{R}\right) 
\widehat{\rho }\left( \underline{R}\right) d^{3}\underline{R}  \nonumber \\
&=&-e\sum_{\underline{Q}}\widehat{\phi }\left( \underline{Q}\right) \int 
\widehat{\rho }\left( {\bf r}\right) e^{i{\bf q\cdot r}}d^{2}{\bf r}\int
\left| f\left( z\right) \right| ^{2}e^{iq_{z}z}dz  \nonumber \\
&=&\frac{ieh_{\mu \nu \lambda }}{\kappa \varepsilon _{0}}\sum_{j\underline{Q}%
}\frac{Q_{\mu }\widehat{{\cal S}}_{\nu \lambda }^{j}\left( \underline{Q}%
\right) }{Q^{2}}\widehat{\rho }_{{\bf -q}}Z\left( q_{z}\right)  \nonumber \\
&=&-\sum_{j\underline{Q}}\Lambda _{j}\left( \underline{Q}\right) Z\left(
q_{z}\right) \widehat{\rho }_{{\bf -q}}\left( \widehat{a}_{j}^{\dagger
}\left( \underline{Q}\right) +\widehat{a}_{j}\left( -\underline{Q}\right)
\right)
\end{eqnarray}
where 
\[
\Lambda _{j}\left( \underline{Q}\right) =\frac{-eh_{\mu \nu \lambda }}{%
2\kappa \varepsilon _{0}}\sqrt{\frac{\hbar }{2N_{0}V\omega _{j}\left( 
\underline{Q}\right) }}\frac{Q_{\mu }\left( \xi _{\nu }^{j}\left( \underline{%
Q}\right) Q_{\lambda }+Q_{\nu }\xi _{\lambda }^{j}\left( \underline{Q}%
\right) \right) }{Q^{2}} 
\]
which, for GaAs reduces to 
\begin{equation}
\Lambda _{j}\left( \underline{Q}\right) =\frac{-eh_{14}}{\kappa \varepsilon
_{0}}\sqrt{\frac{\hbar }{2N_{0}V\omega _{j}\left( \underline{Q}\right) }}%
\frac{Q_{x}Q_{y}\xi _{z}^{j}\left( \underline{Q}\right) +Q_{y}Q_{z}\xi
_{x}^{j}\left( \underline{Q}\right) +Q_{z}Q_{x}\xi _{y}^{j}\left( \underline{%
Q}\right) }{Q^{2}}\qquad .
\end{equation}

In conventional work on phonons in semiconductors it is customary to
regularize the $1/Q$ dependence by including screening of the interaction in
the Thomas-Fermi approximation leading to 
\begin{equation}
\left| \Lambda _{j}\left( \underline{Q}\right) \right| ^{2}\sim \frac{\left|
\gamma _{j}^{pa}\right| ^{2}}{Q+Q_{TF}}
\end{equation}
where $Q_{TF}$ is the usual inverse screening length. We shall not do this
here, partly because there is no coupling to small $Q$ phonons because of
the presence of a gap, but mostly on the grounds that we are directly
employing an approximation (and a very good one in the regime of interest)
for the dynamic structure factor which includes all screening effects. In
practice of course the incompressible nature of the fractional quantum Hall
state means that potentials such as this are hardly screened at all.

Hence we can write the overall electron-phonon coupling Hamiltonian as 
\begin{equation}
=-\sum_{j\underline{Q}}M_{j}\left( \underline{Q}\right) Z\left( q_{z}\right)
\rho _{{\bf -q}}\left( \widehat{a}_{j}^{\dagger }\left( \underline{Q}\right)
+\widehat{a}_{j}\left( -\underline{Q}\right) \right)
\end{equation}
where 
\begin{equation}
M_{j}\left( \underline{Q}\right) =\Lambda _{j}\left( \underline{Q}\right)
-i\Xi _{j}\left( \underline{Q}\right) \qquad .
\end{equation}
If we choose a basis for single-particle electron states $\left\{ \left|
m\right\rangle \right\} $ then we can write 
\begin{eqnarray}
\widehat{\rho }\left( {\bf r}\right) &=&\sum_{mn}\widehat{c}_{m}^{\dagger }%
\widehat{c}_{n}\left\langle m|{\bf r}\right\rangle \left\langle {\bf r|}%
n\right\rangle \\
\widehat{\rho }_{{\bf q}} &=&\sum_{mn}\Gamma _{mn}\left( {\bf q}\right) 
\widehat{c}_{m}^{\dagger }\widehat{c}_{n}
\end{eqnarray}
where 
\begin{equation}
\Gamma _{mn}\left( {\bf q}\right) =\left\langle m\left| e^{i{\bf q\cdot }%
\widehat{{\bf r}}}\right| n\right\rangle
\end{equation}
so that 
\begin{equation}
H_{e\phi }=-\sum_{j\underline{Q},mn}\Gamma _{mn}\left( {\bf q}\right)
M_{j}\left( \underline{Q}\right) \widehat{c}_{m}^{\dagger }\widehat{c}%
_{n}\left( \widehat{a}_{j}^{\dagger }\left( \underline{Q}\right) +\widehat{a}%
_{j}\left( -\underline{Q}\right) \right)
\end{equation}
which bears the usual interpretation of a phonon absorption/emission causing
an electron to scatter from state $n$ to state $m$.

\section{The Isotropic Debye Approximation}

\label{Debye} In the isotropic Debye approximation we neglect the anisotropy
in the speeds of sound and set $\omega _{j}\left( \underline{Q}\right)
=c_{j}Q$, where $c_{1}=c_{LA}$ and $c_{2}=c_{3}=c_{TA}$. The phonon
polarizations become strictly transverse and longitudinal. A suitable set of
orthonormal polarization vectors would be 
\begin{eqnarray*}
\underline{\xi ^{1}}\left( \underline{Q}\right) &=&\underline{Q}/Q=\left(
\sin \theta \cos \phi ,\sin \theta \sin \phi ,\cos \theta \right) \\
\underline{\xi ^{2}}\left( \underline{Q}\right) &=&\left( -\sin \phi ,\cos
\phi ,0\right) \\
\underline{\xi ^{3}}\left( \underline{Q}\right) &=&\left( \cos \theta \cos
\phi ,\cos \theta \sin \phi ,-\sin \theta \right)
\end{eqnarray*}
but any combination of the form 
\begin{eqnarray*}
\underline{\xi ^{TA_{1}}}\left( \underline{Q}\right) &=&\cos \left( \alpha
\right) \underline{\xi ^{2}}\left( \underline{Q}\right) +\sin \left( \alpha
\right) \underline{\xi ^{3}}\left( \underline{Q}\right) \\
\underline{\xi ^{TA_{2}}}\left( \underline{Q}\right) &=&-\sin \left( \alpha
\right) \underline{\xi ^{2}}\left( \underline{Q}\right) +\cos \left( \alpha
\right) \underline{\xi ^{3}}\left( \underline{Q}\right)
\end{eqnarray*}
is equally good. Fortunately in final expressions for energy transfer rates
all dependence on $\alpha $ disappears.

In this approximation the deformation potential coupling has the form 
\[
\Xi _{j}\left( \underline{Q}\right) =\gamma ^{dp}\delta _{j1}\sqrt{Q} 
\]
where 
\[
\gamma ^{dp}=\Xi _{0}\sqrt{\frac{\hbar }{2N_{0}Vc_{j}}} 
\]
so that the electrons only couple to the LA\ phonons via the deformation
potential and the coupling is isotropic. Similarly the piezoelectric
couplings have the form 
\begin{eqnarray*}
\Lambda _{1}\left( \underline{Q}\right) &=&\frac{\gamma _{LA}^{pa}}{\sqrt{Q}}%
A_{1}\left( \theta \right) B_{1}\left( \phi \right) \\
\Lambda _{j}\left( \underline{Q}\right) &=&\frac{\gamma _{TA}^{pa}}{\sqrt{Q}}%
A_{j}\left( \theta \right) B_{j}\left( \phi \right) \qquad j=2,3
\end{eqnarray*}
where $\theta $ and $\phi $ are the polar angles of the $Q$-vector and 
\begin{eqnarray*}
A_{1}\left( \theta \right) &=&3\sin ^{2}\theta \cos \theta \qquad
B_{1}\left( \phi \right) =\frac{1}{2}\sin \left( 2\phi \right) \\
A_{2}\left( \theta \right) &=&\sin \theta \cos \theta \qquad B_{2}\left(
\phi \right) =\cos \left( 2\phi \right) \\
A_{3}\left( \theta \right) &=&\sin \theta \left( 3\sin ^{2}\theta -2\right)
\qquad B_{3}\left( \phi \right) =\frac{1}{2}\sin \left( 2\phi \right) \qquad
.
\end{eqnarray*}

Hence we have, in the isotropic approximation 
\begin{eqnarray*}
\left| M_{1}\left( Q,\theta ,\phi \right) \right| ^{2} &=&\frac{\left(
\gamma _{LA}^{pa}\right) ^{2}}{Q}\left( A_{1}\left( \theta \right)
B_{1}\left( \phi \right) \right) ^{2}+\left( \gamma ^{dp}\right) ^{2}Q \\
\left| M_{j}\left( Q,\theta ,\phi \right) \right| ^{2} &=&\frac{\left(
\gamma _{TA}^{pa}\right) ^{2}}{Q}\left( A_{j}\left( \theta \right)
B_{j}\left( \phi \right) \right) ^{2}\qquad j=2,3\qquad .
\end{eqnarray*}
Since the only separate dependence on the two TA\ modes is in the $A$ and $B$
coefficients we define, using the fact that $A_{j}\left( \theta \right) $
can be written solely as a function of $\sin \theta $, 
\begin{eqnarray*}
\Gamma _{LA}\left( Q,\sin \theta ,\phi \right) &=&\frac{\left( \gamma
_{LA}^{pa}\right) ^{2}}{Q}\left( A_{1}\left( \theta \right) B_{1}\left( \phi
\right) \right) ^{2}+\left( \gamma ^{dp}\right) ^{2}Q \\
\Gamma _{TA}\left( Q,\sin \theta ,\phi \right) &=&\frac{\left( \gamma
_{TA}^{pa}\right) ^{2}}{Q}\left[ \left( A_{2}\left( \theta \right)
B_{2}\left( \phi \right) \right) ^{2}+\left( A_{3}\left( \theta \right)
B_{3}\left( \phi \right) \right) ^{2}\right] \qquad .
\end{eqnarray*}

\section{Energy Transfer Rate}

\label{transferrate} In order to derive an expression for the rate at which
energy is transferred from the phonon pulse to the electron gas we will use
standard linear response theory. We make the usual adiabatic switching-on
assumption and suppose that in the distant past ($t\rightarrow -\infty $)
the electron and phonon systems were uncoupled with the electrons in
equilibrium at temperature $T_{e}$. We will take the initial state of the
phonons to be described by the density matrix 
\begin{equation}
\varrho _{\phi }=\frac{1}{Z_{\phi }}{\cal P}_{\chi }e^{-H_{\phi }/T_{\phi }}%
{\cal P}_{\chi }
\end{equation}
where ${\cal P}_{\chi }$ is a projection operator which excludes states
corresponding to phonons that have not reached the vicinity of the 2DES and $%
H_{\phi }$ is the phonon Hamiltonian.

The initial density matrix of the system is then, 
\begin{equation}
\varrho _{0}=\frac{1}{Z_{e}}e^{-H_{e}/T_{e}}\otimes \frac{1}{Z_{\phi }}{\cal %
P}_{\chi }e^{-H_{\phi }/T_{\phi }}{\cal P}_{\chi }
\end{equation}
where $H_{e}$ is the Hamiltonian for the electron liquid. The time evolution
of the density matrix is governed by the von Neumann equation 
\begin{equation}
i\frac{d\varrho \left( t\right) }{dt}=\left[ H_{e}+H_{\phi }+H_{e\phi
},\varrho \left( t\right) \right] \qquad .
\end{equation}
As usual it is convenient to change to the interaction picture of quantum
mechanics from the usual Schrodinger one by defining 
\begin{eqnarray}
\widetilde{O}\left( t\right) &=&e^{iH_{0}t}Oe^{-iH_{0}t} \\
H_{0} &=&H_{e}+H_{\phi }
\end{eqnarray}
so that 
\begin{equation}
i\frac{d\widetilde{\varrho }\left( t\right) }{dt}=\left[ \widetilde{H}%
_{e\phi }\left( t\right) ,\widetilde{\varrho }\left( t\right) \right]
\end{equation}
which has the iterative solution (to lowest order in the electron phonon
coupling) 
\begin{equation}
\widetilde{\varrho }\left( t\right) =\varrho _{0}-i\int_{-\infty }^{t}\left[ 
\widetilde{H}_{e\phi }\left( t^{\prime }\right) ,\varrho _{0}\right] e^{\eta
t^{\prime }}dt^{\prime }
\end{equation}
where $\eta $ is the adiabatic switch-on rate. The rate of energy transfer
to the electron gas is 
\begin{eqnarray}
P &=&\frac{d}{dt}\left\langle H_{e}\right\rangle =-\frac{d}{dt}\left\langle
H_{\phi }\right\rangle  \nonumber \\
&=&i\mathop{\rm tr}\left\{ \widetilde{\varrho }\left( t\right) \left[ 
\widetilde{H_{\phi }}\left( t\right) ,\widetilde{H_{e\phi }}\left( t\right) %
\right] \right\}  \nonumber \\
&=&\int_{-\infty }^{t}\mathop{\rm tr}\left\{ \varrho _{0}\left[ \left[ 
\widetilde{H_{\phi }}\left( t\right) ,\widetilde{H_{e\phi }}\left( t\right) %
\right] ,\widetilde{H}_{e\phi }\left( t^{\prime }\right) \right] \right\}
e^{\eta t^{\prime }}dt^{\prime }\qquad .
\end{eqnarray}
Now 
\begin{equation}
\left[ H_{\phi },H_{e\phi }\right] =-\sum_{j\underline{Q}}\omega _{j}\left( 
\underline{Q}\right) \left( M_{j}\left( \underline{Q}\right) \rho _{{\bf -q}%
}a_{j}^{\dagger }\left( \underline{Q}\right) -M_{j}\left( -\underline{Q}%
\right) \rho _{{\bf q}}a_{j}\left( \underline{Q}\right) \right) \qquad ,
\end{equation}
\begin{equation}
\left\langle \widetilde{a}_{j}^{\dagger }\left( \underline{Q},t\right) 
\widetilde{a}_{j^{\prime }}\left( \underline{Q}^{\prime },t^{\prime }\right)
\right\rangle =\delta _{jj^{\prime }}\delta _{\underline{Q},\underline{%
Q^{\prime }}}n_{j}\left( \underline{Q}\right) e^{i\omega _{j}\left( 
\underline{Q}\right) \left( t-t^{\prime }\right) }
\end{equation}
and 
\begin{equation}
M_{j}\left( -\underline{Q}\right) =M_{j}^{\ast }\left( \underline{Q}\right)
\qquad .
\end{equation}
Hence, after some rather tedious algebra and setting $\eta =0$, we find 
\begin{equation}
P=2\pi \overline{\rho }\Omega \sum_{j\underline{Q}}\omega _{j}\left( 
\underline{Q}\right) \left| M_{j}\left( \underline{Q}\right) \right|
^{2}\left\{ n_{j}\left( \underline{Q}\right) S\left( {\bf q},\omega
_{j}\left( \underline{Q}\right) \right) -\left( 1+n_{j}\left( \underline{Q}%
\right) \right) S\left( -{\bf q},-\omega _{j}\left( \underline{Q}\right)
\right) \right\}
\end{equation}
where 
\begin{equation}
S\left( {\bf q},\omega \right) =\frac{1}{2\pi \overline{\rho }\Omega }\int
\left\langle \rho _{-{\bf q}}\left( t\right) \rho _{{\bf q}}\left( 0\right)
\right\rangle e^{i\omega t}dt
\end{equation}
is the dynamic structure factor. As usual (see next appendix) \cite{Quantum
liquids} 
\begin{equation}
S\left( {\bf -q},-\omega \right) =e^{-\omega /T_{e}}S\left( {\bf q},\omega
\right)
\end{equation}
so that 
\begin{equation}
P=2\pi \overline{\rho }\Omega \sum_{j\underline{Q}}\omega _{j}\left( 
\underline{Q}\right) \left| M_{j}\left( \underline{Q}\right) \right|
^{2}S\left( {\bf q},\omega _{j}\left( \underline{Q}\right) \right) \left\{
n_{j}\left( \underline{Q}\right) -e^{-\omega _{j}\left( \underline{Q}\right)
/T_{e}}\left( 1+n_{j}\left( \underline{Q}\right) \right) \right\} \qquad .
\end{equation}

\section{Spectral Analysis of the Dynamic Structure Factor}

\label{Structure} Let us consider the exact energy and electron number
eigenstates of the electronic system $\left\{ \left| \alpha \right\rangle
\right\} $. The dynamic structure factor can then be written 
\begin{eqnarray}
S\left( {\bf q},\omega \right) &=&\frac{1}{2\pi \overline{\rho }\Omega }\int 
\frac{1}{Z_{e}}\sum_{\alpha \gamma }\left\langle \alpha \right|
e^{-H_{e}/T_{e}}e^{iH_{e}t}\rho _{-{\bf q}}e^{-iH_{e}t}\left| \gamma
\right\rangle \left\langle \gamma \right| \rho _{{\bf q}}\left| \alpha
\right\rangle e^{i\omega t}dt  \nonumber \\
&=&\frac{1}{\overline{\rho }\Omega }\frac{1}{Z_{e}}\sum_{\alpha \gamma
}\left| \left\langle \gamma \right| \rho _{{\bf q}}\left| \alpha
\right\rangle \right| ^{2}e^{-E_{\alpha }/T_{e}}\delta \left( \omega -\left(
E_{\gamma }-E_{\alpha }\right) \right)
\end{eqnarray}
where $Z_{e}=\sum_{\alpha }e^{-E_{\alpha }/T_{e}}$ is the partition function
for the electronic system. Now 
\begin{equation}
S\left( -{\bf q},-\omega \right) =\frac{1}{\overline{\rho }\Omega }\frac{1}{%
Z_{e}}\sum_{\alpha \gamma }\left| \left\langle \gamma \right| \rho _{-{\bf q}%
}\left| \alpha \right\rangle \right| ^{2}e^{-\left( E_{\alpha }-\mu
N_{\alpha }\right) /T_{e}}\delta \left( -\omega -\left( E_{\gamma
}-E_{\alpha }\right) \right)
\end{equation}
interchanging the dummy variables gives 
\begin{eqnarray}
S\left( -{\bf q},-\omega \right) &=&\frac{1}{\overline{\rho }\Omega }\frac{1%
}{Z_{e}}\sum_{\alpha \gamma }\left| \left\langle \alpha \right| \rho _{-{\bf %
q}}\left| \gamma \right\rangle \right| ^{2}e^{-E_{\gamma }/T_{e}}\delta
\left( -\omega -\left( E_{\alpha }-E_{\gamma }\right) \right)  \nonumber \\
&=&\frac{1}{\overline{\rho }\Omega }\frac{1}{Z_{e}}\sum_{\alpha \gamma
}\left| \left\langle \gamma \right| \rho _{{\bf q}}\left| \alpha
\right\rangle \right| ^{2}e^{-\left( E_{\alpha }+\omega \right)
/T_{e}}\delta \left( \omega -\left( E_{\gamma }-E_{\alpha }\right) \right) 
\nonumber \\
&=&e^{-\omega /T_{e}}S\left( {\bf q},\omega \right)
\end{eqnarray}
as required in appendix \ref{transferrate}.

Now let us write 
\begin{eqnarray}
Z_{e} &=&\sum_{\alpha }e^{-E_{\alpha }/T_{e}}  \nonumber \\
&=&e^{-E_{0}/T_{e}}\left( 1+\sum_{\alpha \neq 0}e^{-\left( E_{\alpha
}-E_{0}\right) /T_{e}}\right)
\end{eqnarray}
where $E_{0}$ is the ground state energy. We know that $E_{\alpha
}-E_{0}>\Delta ^{\ast }$ $\forall \alpha >0$ so that 
\begin{equation}
Z\sim e^{-E_{0}/T_{e}}\left( 1+O\left( e^{-\Delta ^{\ast }/T_{e}}\right)
\right) \qquad T_{e}\rightarrow 0\qquad .
\end{equation}
Similarly 
\begin{eqnarray}
Z_{e}S\left( {\bf q},\omega \right) &=&\frac{1}{\overline{\rho }\Omega }%
\sum_{\alpha \gamma }\left| \left\langle \gamma \right| \rho _{{\bf q}%
}\left| \alpha \right\rangle \right| ^{2}e^{-E_{\alpha }/T_{e}}\delta \left(
\omega -\left( E_{\gamma }-E_{\alpha }\right) \right)  \nonumber \\
&=&e^{-E_{0}/T_{e}}\left\{ \frac{1}{\overline{\rho }\Omega }\sum_{\gamma
}\left| \left\langle \gamma \right| \rho _{{\bf q}}\left| 0\right\rangle
\right| ^{2}\delta \left( \omega -\left( E_{\gamma }-E_{0}\right) \right)
\right.  \nonumber \\
&&+\left. \frac{1}{\overline{\rho }\Omega }\sum_{\alpha \neq 0}e^{-\left(
E_{\alpha }-E_{0}\right) /T_{e}}\sum_{\gamma }\left| \left\langle \gamma
\right| \rho _{{\bf q}}\left| \alpha \right\rangle \right| ^{2}\delta \left(
\omega -\left( E_{\gamma }-E_{\alpha }\right) \right) \right\}  \nonumber \\
S\left( {\bf q},\omega \right) &\sim &\frac{1}{\overline{\rho }\Omega }%
\sum_{\gamma }\left| \left\langle \gamma \right| \rho _{{\bf q}}\left|
0\right\rangle \right| ^{2}\delta \left( \omega -\left( E_{\gamma
}-E_{0}\right) \right) +O\left( e^{-\Delta ^{\ast }/T_{e}}\right) \qquad
T_{e}\rightarrow 0\qquad .
\end{eqnarray}
Where the leading term is simply the zero temperature structure factor.

\section{Simplification of $P$ Integral}

\label{integral} We have an integral of the form 
\begin{equation}
P_{j}=\int_{0}^{\infty }dQQ^{2}\int_{0}^{\pi }d\theta \sin \theta F\left(
c_{j}Q,Q\sin \theta ,Q\cos \theta \right)
\end{equation}
We introduce new variables $\omega =c_{j}Q$ \ and $q=Q\sin \theta $. Now 
\begin{eqnarray}
\left| \frac{\partial \left( \omega ,q\right) }{\partial \left( Q,\theta
\right) }\right| &=&\left| 
\begin{array}{cc}
c_{j} & 0 \\ 
\sin \theta & Q\cos
\end{array}
\right|  \nonumber \\
&=&c_{j}Q\cos \theta  \nonumber \\
&=&\sqrt{\omega ^{2}-c_{j}^{2}q^{2}}
\end{eqnarray}
whence 
\begin{equation}
Q^{2}\sin \theta dQd\theta =\frac{\omega q}{c_{j}\sqrt{\omega
^{2}-c_{j}^{2}q^{2}}}d\omega dq
\end{equation}
so that 
\begin{equation}
P_{j}=\int_{0}^{\infty }d\omega \int_{0}^{\infty }dqF\left( \omega ,q,\frac{%
\sqrt{\omega ^{2}-c_{j}^{2}q^{2}}}{c_{j}}\right) \frac{\omega q\vartheta
\left( \omega -c_{j}q\right) }{c_{j}\sqrt{\omega ^{2}-c_{j}^{2}q^{2}}}
\end{equation}
which leads to the formula quoted in the text.

\begin{figure}[tbp]
\caption{Schematic view of the devices used in phonon spectroscopy
experiments showing the 2des connected by long lines to 3d contact pads and
two heaters, H1 and H2 mounted on the back face. The inset shows the 2d and
3d wavevector conventions used.}
\label{schematic}
\end{figure}

\begin{figure}[tbp]
\caption{Schematic graph illustrating the dispersion relation for
magnetorotons with the dispersion curves for $\protect\theta =30^{\circ }$
phonons.}
\label{fig2}
\end{figure}

\begin{figure}[tbp]
\caption{Extreme LA lines (corresponding to $\protect\theta _{\min }$ and $%
\protect\theta _{\max }$) superposed on the magnetoroton dispersion
corresponding to $\Delta ^{\ast }=0.05{\cal E}_{c}$.}
\label{fig3a}
\end{figure}

\begin{figure}[tbp]
\caption{Extreme TA lines (corresponding to $\protect\theta _{\min }$ and $%
\protect\theta _{\max }$) superposed on the magnetoroton dispersion
corresponding to $\Delta ^{\ast }=0.05{\cal E}_{c}$.}
\label{fig3b}
\end{figure}

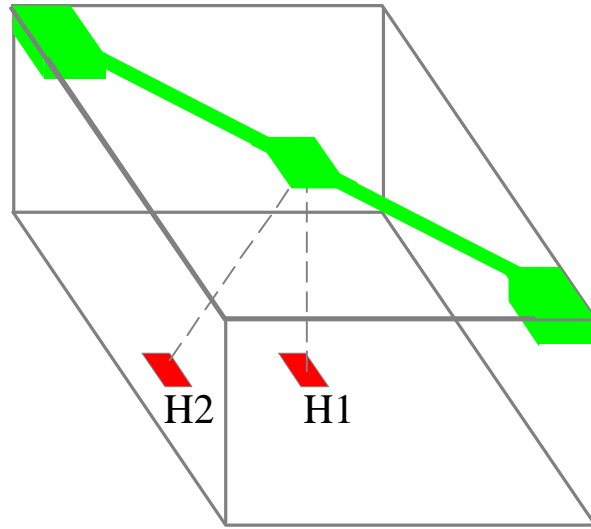
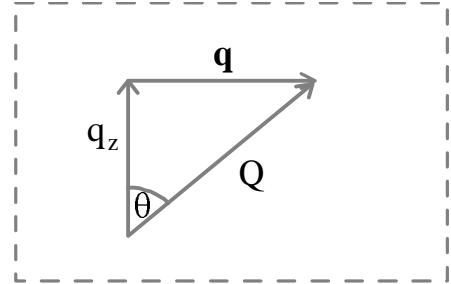
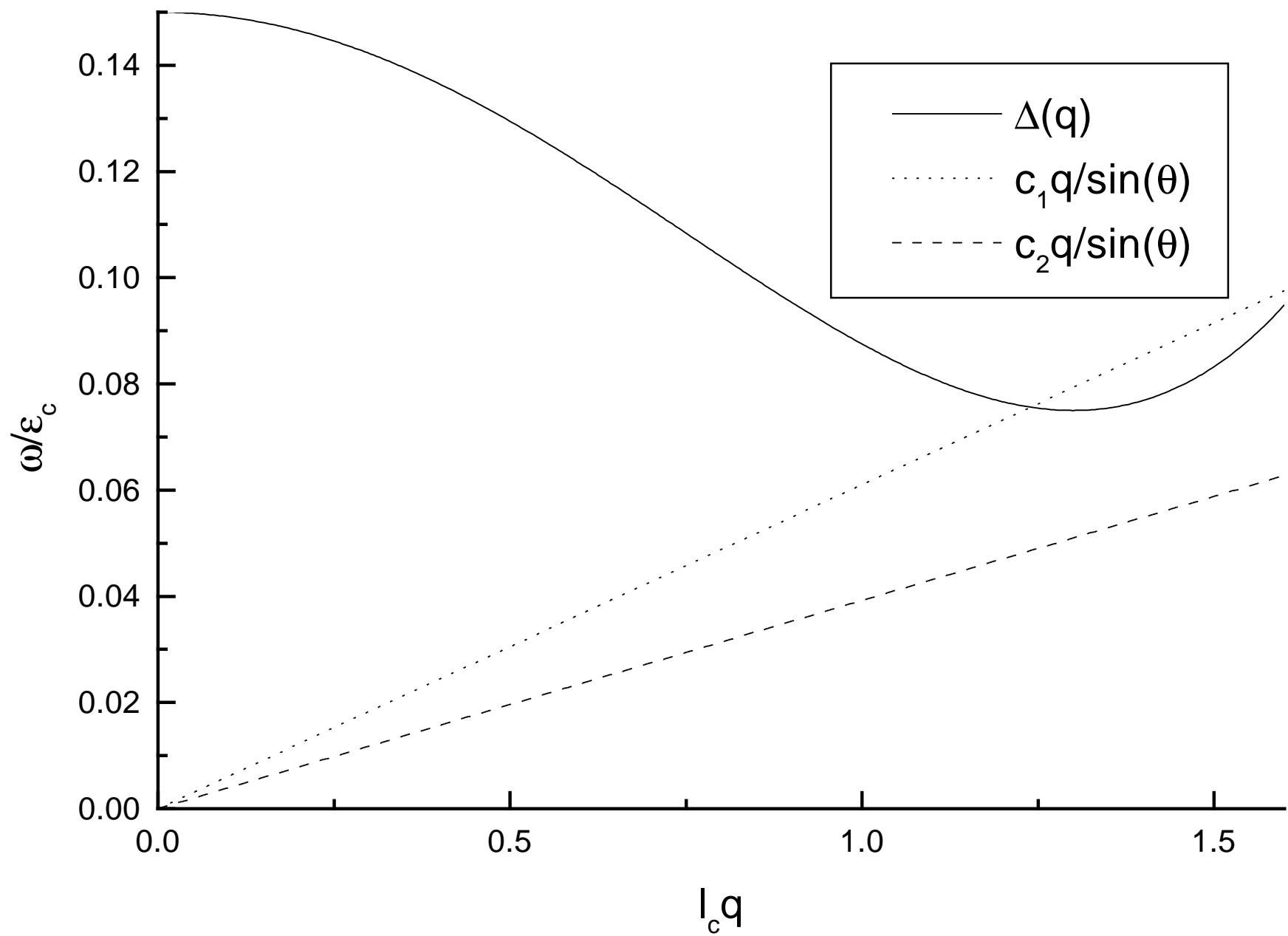
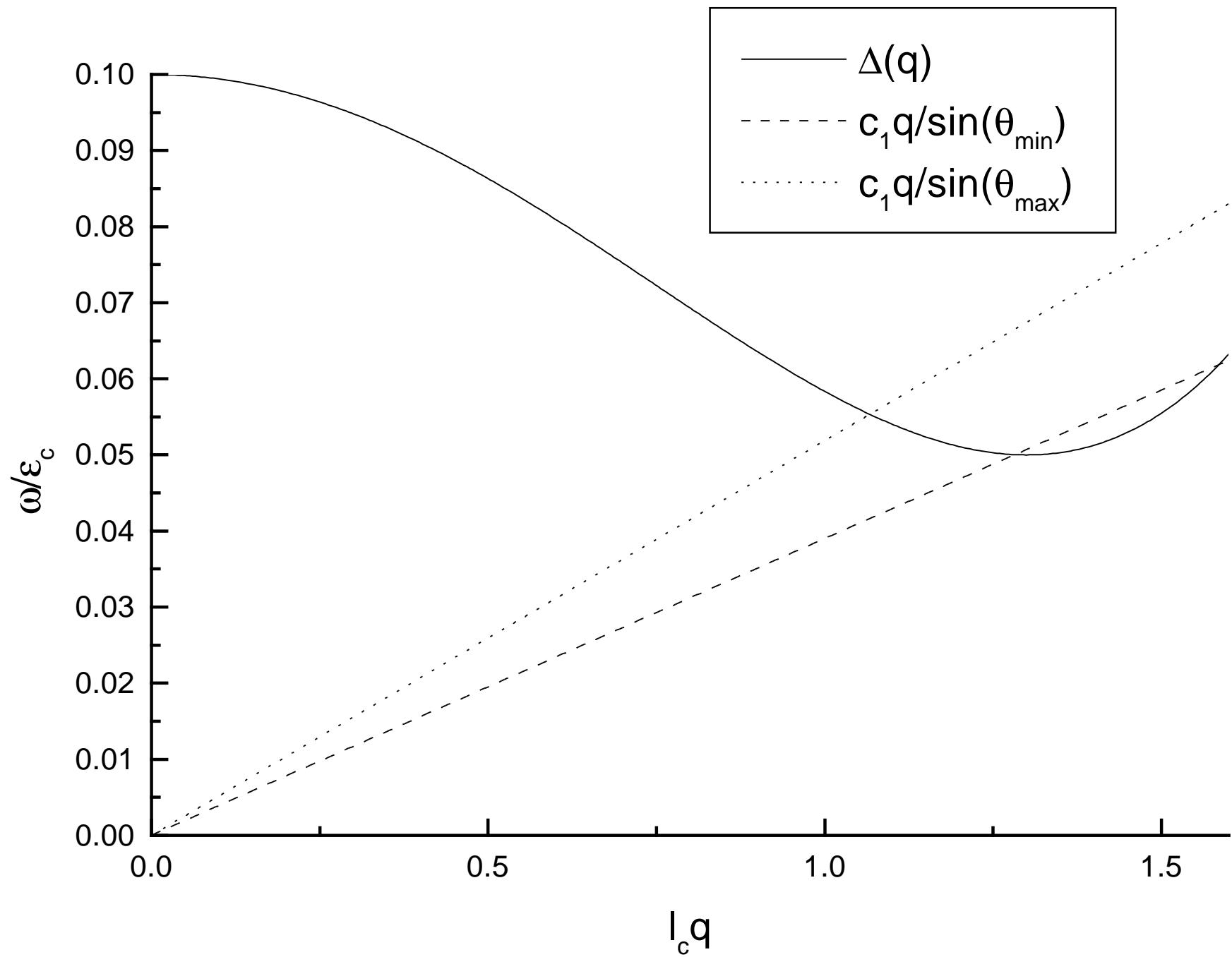
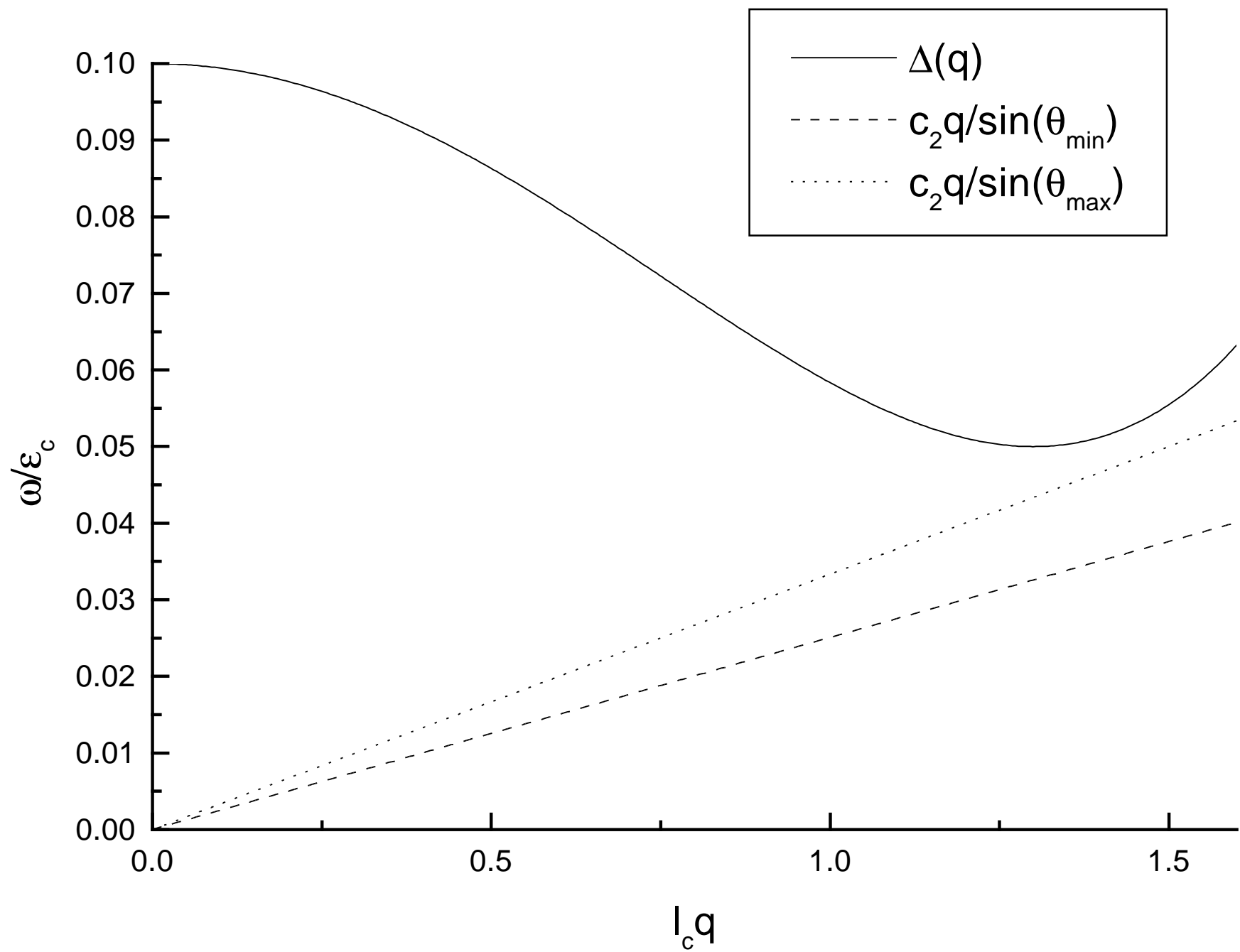
\end{document}